\documentclass[aps,twocolumn,showpacs,floatfix,nofootinbib,10pt]{revtex4-1}
\usepackage{amssymb,amsmath}
\usepackage{graphicx,float,color}
\usepackage{indentfirst}
\newcommand{\be}{\begin{equation}}
\newcommand{\ee}{\end{equation}}

\newcommand{\ba}{\begin{eqnarray}}
\newcommand{\ea}{\end{eqnarray}}

\begin{document}
%
%
%\title{The minimal geometric deformation approach extended:\\ a new exact solution in the braneworld}
\title{Extending the geometric deformation: New black hole solutions}
\author{J. Ovalle}
\email{jovalle@ictp.it\\jovalle@usb.ve}
\affiliation{Abdus Salam International Center for Theoretical Physics (ICTP), Strada Costiera 11, Trieste 34014, Italy
\\
Dipartimento di Fisica e Astronomia, Alma Mater Universit\`a di Bologna, 40126 Bologna, Italy
\\
Departamento de F\'isica, Universidad Sim\'on Bol\'ivar, Apartado 89000, Caracas 1080A,
Venezuela}

\begin{abstract}
By using the extension of the Minimal Geometric Deformation approach, recently developed to investigate the exterior spacetime of a self-gravitating system in 
the Braneworld, we identified a master solution for the deformation undergone by the radial metric component when time deformations are produced by bulk gravitons. 
A specific form for the temporal deformation is used to generate a new exterior solution with a tidal charge $Q$. The main feature of this solution is the presence 
of higher-order terms in the tidal charge, thus generalizing the well known tidally charged solution. The horizon of the black hole lies inside 
the Schwarzschild radius, $h<r_s=2\,{\cal M}$, indicating that extra-dimensional effects weaken the gravitational field.\footnote{Based on the contributed lecture given at the 9th Alexander Friedmann International Seminar, June 21-27, 2015, 
St. Petersburg, Russia.}
\end{abstract}\maketitle
%\keywords{braneworlds, black holes, extra-dimensional gravity}
%\arxivnumber{arXiv:***}
\flushbottom
%
%\pacs{11.25.-w, 04.50.-h, 04.50.Gh}
%
%
\section{Introduction}

\par
General Relativity (GR), in its hundred years of existence, has proved to be a successful and well tested theory for gravitation. 
Along with the Standard Model (SM) of particle physics, both theories represent the fundamental structure of modern physics. 
However, it is fair to say that GR presents serious problems in the proper description of gravity at both, very small scales and beyond 
the Solar System scales, namely, its quantum formulation and the dark matter problem. Besides needing dark energy to explain the accelerating universe. 
This is the main motivation for the search for new theories beyond GR which could explain these fundamental issues. Among them we may mention a large number of 
extra-dimensional theories and high curvature gravity theories, most of them inspired or a direct consequence of Superstring theory. We  also have Galileon theories, 
$f(R)$ gravity theories, Scalar-tensor theories, Massive Gravity, New Massive Gravity, Topologically Massive Gravity, Chern-Simons theories, 
Higher Spin Gravity theories, Horava-Lifshitz Gravity, etc. (see for instance Refs.~\cite{Ber}--\cite{horava}). Despite this great effort, the fundamental issues associated with the gravitational 
interaction remain unresolved.
\par
Among extra-dimensional theories we have the Randall Sundrum Braneworld (BW) \cite{lisa1,lisa2} which not only explain the hierarchy of fundamental 
interactions \cite{ADD,ADD1} but also is formulated in a non-trivial bulk, which
makes it more attractive to explore gravity at high energies as well as generalizations of four-dimensional GR. Because of this, its study and impact on GR is fully 
justified (see, for instance, Refs.~\cite{shi}--\cite{RRR}). 
In this context, by considering the impact of the BW on self-gravitating system, the Minimal Geometric Deformation (MGD) was developed \cite{jo1, GD100}. 
This approach  
has proven to be useful, among other things, to derive exact and physically
acceptable  solutions  for  spherically  symmetric  and
non-uniform stellar distributions \cite{jo2}-\cite{jo5} as well, to  express  the  tidal  charge  in  the  metric  found  in
Ref.~\cite{qmetric} in terms of the  Arnowitt-Deser-Misner (ADM) mass \cite{jo6}, to study microscopic black holes \cite{jo7}, 
to clarify the role of exterior Weyl stresses  acting  on  compact  stellar  
distributions \cite{jo8,jo9}, as well as to extend the concept of variable tension introduced in Ref.~\cite{ger1} by analyzing the
black strings solution into the extra dimension \cite{jo10},
to prove, contrary to previously established, the existence of Schwarzschild exterior for a spherically symmetric BW self-gravitanting system made of
regular matter \cite{jo11}, and to derive bounds on extra-dimensional parameters from the observational results of the classical 
tests of GR in the Solar system \cite{jo12}.

\par
In a recent work \cite{MGDextended}, by studying the exterior spacetime $r>R$ associated to a BW self-gravitating system of radius $R$ (where the GR vacuum $r>R$ is 
filled with a Weyl fluid of extra-dimensional origin), the MGD approach was successfully extended to the most general case for spherically symmetric distributions, namely, 
when both gravitational potential $g_{tt}$ and $g_{rr}$ are deformed by bulk gravitons. It was shown than the deformation undergone by the temporal metric component produces part of the 
deformation undergone by the radial metric component. In the present work, by using the extension of the MGD approach and a time deformation parameter, 
a master solution for the radial geometric deformation associated to the time deformation is identified. Hence by choosing a specific form of the time deformation, 
a new black hole solution is identified, showing that extra-dimensional effects weaken the gravitational field.
\par
In the next section, we shall present the standard static BW equations for the exterior $r>R$ of a spherically symmetric self-gravitating system, 
which, contrary to GR, is filled with a Weyl fluid of extra-dimensional origen. In section 2 a brief summary of the 
generalization of the MGD approach is presented. In section 3 we shall identify a master solution for the radial metric component 
associated to the time deformation, and a new black hole solution is described in detail
in section 4. Finally we summarize our work in section 5.  

\section{Field Equations}	

\par
In the generalised Randall-Sundrum scenario, gravity acts in the five dimensional bulk $(3+1+1)$ 
and modifies the gravitational dynamics in the observable four-dimensional world.
In the vacuum $T_{\mu \nu}=0$ the effective four-dimensional Einstein equations are written as\footnote{We use
$k^{2}=8\,\pi \,G$, where $G$ is the 4-dimensional Newton constant, and $\Lambda$ is 
the 4-dimensional cosmological constant.} 
\begin{equation}
G_{\mu \nu }
=
-k^{2}\,T_{\mu \nu }^{\mathrm{eff}}-\Lambda \,g_{\mu \nu }
\ ,
\label{4Dein}
\end{equation}
where the effective energy-momentum tensor
\begin{equation}
T_{\mu \nu }^{\mathrm{eff}}
=
\frac{1}{8\pi }\,\mathcal{E}_{\mu \nu }
+\frac{4}{\sigma }\mathcal{F}_{\mu \nu }
\ ,
\label{tot}
\end{equation}
encodes all fields with an extra-dimensional origen. Here $\sigma$ is the brane tension and 
\begin{equation}
k^{2}\,\mathcal{E}_{\mu \nu }
=
\frac{6}{\sigma }\left[
\mathcal{U}\left(u_{\mu }\,u_{\nu }+\frac{1}{3}\,h_{\mu \nu }\right)
+\mathcal{P}_{\mu \nu }\right]\ ,
\vspace{0.0cm}\end{equation}
where $\mathcal{U}$ is the bulk Weyl scalar, $\mathcal{P}_{\mu \nu }$ the stress tensor and $\mathcal{F}_{\mu \nu }$ contains contributions
from all non-standard model fields possibly living in the bulk. 
For simplicity, we shall assume $\mathcal{F}_{\mu \nu }=0$ and $\Lambda =0$ throughout the paper.
\par
In this extra-dimensional context, the exterior $r>R$ of a spherically symmetric self-gravitating system of radius $R$ will be filled with a Weyl fluid 
of effective density $\mathcal{U}$ and anisotropy $\mathcal{P}_{\mu \nu}$. This fluid has interesting phenomenological consequences, and indeed may be used to explain 
some issues associated with the gravitational interaction beyond the Solar System scales \cite{tiberio1}-\cite{tiberio3}.
\par
By using the Schwarzschild-like coordinates of the metric 
\begin{equation}
ds^{2}
=
e^{\nu(r)}\,dt^{2}
-e^{\lambda(r)}\,dr^{2}
-r^{2}\left( d\theta^{2}+\sin ^{2}\theta \,d\phi ^{2}\right)
\ ,
\label{metric}
\end{equation}
the field equations~\eqref{4Dein} are written as 
%\begin{widetext}
\begin{eqnarray}
&&
k^{2}\left(
\strut \displaystyle\frac{1}{\sigma }\frac{6}{k^{4}}\,\mathcal{U} \right)
=
\strut \displaystyle\frac{1}{r^{2}}
-e^{-\lambda }\left( \frac{1}{r^{2}}-\frac{\lambda ^{\prime }}{r}\right)\ ,
\label{ec1}
\\
&&
\notag
\\
&&
k^{2}\strut \displaystyle\left(\frac{1}{\sigma }\frac{2}{k^{4}}\,\mathcal{U}
+\frac{4}{k^{4}}\frac{\mathcal{P}}{\sigma }\right)
=
-\frac{1}{r^{2}}+e^{-\lambda }\left( \frac{1}{r^{2}}+\frac{\nu ^{\prime }}{r}\right)\ ,
\label{ec2}
\\
&&
\notag
\\
&&
k^{2}\strut \displaystyle\left(\frac{1}{\sigma }\frac{2}{k^{4}}\mathcal{U} -\frac{2}{k^{4}}\frac{\mathcal{P}}{\sigma }\right)
=
\frac{1}{4}e^{-\lambda }\left[ 2\,\nu ^{\prime \prime}
+\nu ^{\prime 2}-\lambda ^{\prime }\,\nu ^{\prime }+2\,\frac{\nu ^{\prime}
-\lambda ^{\prime }}{r}\right]
\ ,
\label{ec3}
\end{eqnarray}
%\end{widetext}
with primes denoting derivatives with respect to $r$. We also have the conservation equation
\begin{equation}
\label{con}
\nabla^{\mu}{\cal E}_{\mu\nu}=0\ ,
\end{equation}
which holds when field equations~\eqref{ec1}-\eqref{ec3} are satisfied, and the null-fluid condition  
\begin{eqnarray}
\label{R2}
R_\mu^{\ \mu}
&=&
e^{-\lambda}\left(\nu''+\frac{\nu'^2}{2}+2\,\frac{\nu'}{r}
+\frac{2}{r^2}\right)
\nonumber
\\
&&
-\lambda'\,e^{-\lambda}\left(\frac{\nu'}{2}+\frac{2}{r}\right)
-\frac{2}{r^2}
=0\ ,
\end{eqnarray}
which is the same than the vacuum condition in GR. In this case this expression comes from the 
condition ${\cal E}_\mu^{\,\,\mu}=0$ for $r>R$ and it is nothing but a linear combination of field equations~\eqref{ec1}-\eqref{ec3}. 
From these expressions we identify 
the density ${\cal U}$, effective radial pressure $\tilde{p}_{r}
$ and effective tangential pressure $\tilde{p}_{t}$. The effective pressures are given by 
\begin{equation}
\tilde{p}_{r}\,=\left(\frac{{\cal U}}{3}+\frac{2{\cal P}}{3} \right)\ ,  \label{efecprera}
\end{equation}%
\begin{equation}
\tilde{p}_{t}\,=\,\left(\frac{{\cal U}}{3}-\frac{{\cal P}}{3}\right)\ , \label{efecpretan}
\end{equation}%
clearly illustrating the anisotropy of the exterior $r>R$, that is 
\begin{equation}
\Pi \equiv 
\tilde{p}_{r}-\tilde{p}_{t}
=
{\cal P}
\ .
\end{equation}%
We shall see that the fields ${\cal U}$ and ${\cal P}$ involves extra-dimensional effects with a big impact on stellar systems. 
Next a summary of the generalization of the MGD approach is presented.

\section{The Extension of the Minimal Geometric Deformation}

%
%
%\setcounter{equation}{0}
%\ref{MGDE}
%
%
%
The components of the Weyl fluid ${\cal U}(r)$ and ${\cal P}(r)$ filling the exterior $r>R$ and the geometric
functions $\nu(r)$ and $\lambda(r)$ must satisfy the field equations~\eqref{ec1}-\eqref{ec3}. 
Hence it is necessary to provide a condition to close the system. Before imposing any constraint, let us start by using 
the null-fluid condition in Eq.~\eqref{R2} along with the 
extended MGD, which considers the deformation undergone by the Schwarzschild solution as
\begin{equation}
 \label{f}
 e^{-\lambda(r)} = 1-\frac{2\,M}{r}+f(r)
\end{equation}
and 
\begin{equation}
\label{def4}
\nu(r)
=
\nu_s+h(r)
\ ,
\end{equation}
where $\nu_s$ is given by the Schwarzschild expression
\begin{equation}
e^{\nu_s}
=
1-\frac{2\,{M}}{r}
\ ,
\label{Schw00}
\end{equation}
with $f(r)$ the geometric deformation of the radial metric component and $h(r)$ the time deformation. By using the expressions 
in Eq.~\eqref{f} and Eq.~\eqref{def4} in the null-fluid condition given in Eq.~\eqref{R2}, we obtain the following condition for the exterior $r>R$, 
\begin{equation}
 {\left(\frac{\nu'}{2}+\frac{2}{r}\right)}\,f'
+{\left(\nu''+\frac{{\nu'}^2}{2}+\frac{2\nu'}{r}+\frac{2}{r^2}\right)}\,f
+F(h)
=
0
\ ,
\label{defor}
\end{equation}
whose formal solution is given by
\begin{equation}
\label{genvacsol}
f(r)
=
e^{-I(r,r_0)}
\left(
\beta
-\int_R^r\frac{e^{I(x,R)}\,F(h)}{\frac{\nu'}{2}+\frac{2}{x}}
dx
\right)\ ,
\end{equation}
where $\beta$ is an integration constant and the exponent $I=I(r,r_0)$ is given by 
\begin{equation}
\label{I}
I(r,r_0)
\equiv
\int^r_{r_0}
\frac{\left(\nu''+\frac{{\nu'}^2}{2}+\frac{2\nu'}{x}+\frac{2}{x^2}\right)}
{\left(\frac{\nu'}{2}+\frac{2}{x}\right)}\,dx
\ ,
\end{equation}
where $r_0$ is a convenient referential value and with the $F(h)$ function
\begin{equation}
F(h)
=
\mu'\,\frac{h'}{2}+\mu\left(h''+\nu_s'\,h'+\frac{h'^2}{2}+2\,\frac{h'}{r}\right)
\ .
\label{F}
\end{equation}
The exterior deformed radial metric component is finally expressed as
\begin{eqnarray}
e^{-\lambda(r)}
&=&
1-\frac{2\,{M}}{r}
\nonumber
\\
&&
+
\underbrace{e^{-I(r,R)}
\left(\beta -\int_R^r\frac{e^{I(x,R)}\,F(h)}{\frac{\nu'}{2}+\frac{2}{x}}\,
dx\right)}_{\rm Geometric\ deformation}
\ .
\label{genvacsol2}
\end{eqnarray}
Finally, the components of the Weyl fluid ${\cal U}$ and ${\cal P}$ are expressed in terms of the radial
deformation $f(r)$ and time deformation $h(r)$ as
\begin{equation}
 \label{UU}
{\cal U}=-\frac{f}{r^2}-\frac{f'}{r}\ ,
\end{equation}
\begin{equation}
 \label{P}
 2\,{\cal P}=\frac{4\,f}{r^2}+\frac{1}{r}\left[f'+3f\nu'_{s}+3\,e^{\nu_s}h'+3fh'\right]\ ,
\end{equation}
where the expressions in Eqs.~\eqref{def4} and~\eqref{genvacsol2} have been used 
in the field equations~\eqref{ec1} and~\eqref{ec2}. Therefore, according to Eq.~\eqref{genvacsol2}, given a time deformation $h=h(r)$ it will induce part 
of the radial deformation $f=f(r)$ undergone by the radial metric component. It is worth noting that a vanishing time deformation $h=0$ will produce $F=0$, 
in consequence the radial geometric deformation will be minimal.
For the Schwarzschild geometry, this procedure yields the deformed exterior
solution previously studied in Ref.~\cite{MGDextended} (we note a constant $h$
also produces $F=0$, and corresponds to a time transformation 
$dT = e^{h/2}\,dt$).

\section{A New Solution}

%\ref{BH}

It would be interesting to consider vacuum solutions by using the deformation parameter $k$ introduced in Ref.~\cite{MGDextended}, namely 
\begin{equation}
\label{g00def}
 e^{\nu} = \left(1-\frac{{2 M}}{r}\right)^{1+k}\ ,
\end{equation}
this correspond to a temporal deformation given by
\begin{equation}
\label{tdef}
 h (r) = k\,ln\left(1-\frac{{2 M}}{r}\right)\ ,
\end{equation}
where $M$ is a free parameter related with the ADM mass ${\cal M}$ of the self-gravitating system. 
This temporal deformation $h(r)$ induces a deformation in the radial metric component, as clearly is shown in the expression (\ref{genvacsol2}),  
in consequence we will have a modification of the Schwarzschild solution in both gravitational potentials. 
\par
Now when Eq. (\ref{g00def}) is used in Eq. (\ref{genvacsol2}) the deformed radial metric component is given in an exact form by
\begin{eqnarray}
\label{master}
e^{-\lambda(r)}
&=&1-2\,M/r+\nonumber \\
&&\frac{1}{r^k}(1-2\,M/r)^{(1-k)}\left[(k-3)M+2r\right]^{\frac{3-5k}{k-3}}\left(\frac{-2M}{r-2M}\right)^k\left(1+\frac{2r}{(k-3)M}\right)^{\frac{4k}{k-3}}
\nonumber \\ 
&&\left[\beta\,r^{\frac{k(k+1)}{k-3}}\left(1-\frac{r}{2\,M}\right)^k\left(1+\frac{2r}{(k-3)M}\right)^{\frac{4k}{k-3}}\right.
\nonumber \\ &&
\left. -(k-3)M\,r^k(1-2\,M/r)\left[(k-3)M+2r\right]^{\frac{4k}{k-3}}\,F(r,k)\right]\ ,\nonumber \\
\end{eqnarray}
where $F(r,k)$ is the Appell hypergeometric function of two variables, given by
\begin{equation}
 F(r,k)=ApellF1[a, b_1, b_2, c; x, y]\ ,
\end{equation}
where
\begin{eqnarray}
 &&a  = \frac{k(k+1)}{3-k};\, \, b_1  = 1-k;\,\, b_2  = \frac{4\,k}{3-k};\,\,c = \frac{3 + k^2}{ 3 - k}; \nonumber \\
 &&x = \frac{r}{2\,M};\,\,\,y  = \frac{2\,r}{(3-k)M}\ .\nonumber \\
\end{eqnarray}
\par
The expression in Eq.~\eqref{master} represents a kind of master equation associated with the time deformation in Eq.~\eqref{tdef}, in consequence 
a family of exact configurations may be generated by using the ``deformation parameter'' $k$ shown in 
the expression (\ref{g00def}). The simplest one is the minimal geometric deformation 
associated to the Schwarzschild solution, which correspond to $k=0$, namely, no time deformation, yielding to
\begin{equation}
 e^{-\lambda}\,=\,\left(1-\frac{2\,M}{r}\right)\left(1+\frac{\beta}{1-\frac{3\,M}{2\,r}}\frac{\ell_0}{r}\right)\ ,
\end{equation}
with
\begin{equation}
 \ell_0\equiv\,R\,\frac{(1-\frac{3M}{2R})}{(1-\frac{2M}{R})}\ .
\end{equation}
Of course when $\beta = 0$ we regain the well known Schwarzschild solution. A more interesting case, still not considered, is 
the one where $k=4$, which represents a solution with high order deformation in terms of a tidal charge $Q$
\par
\begin{figure}[pb]
\begin{center}\includegraphics[width=3.2in]{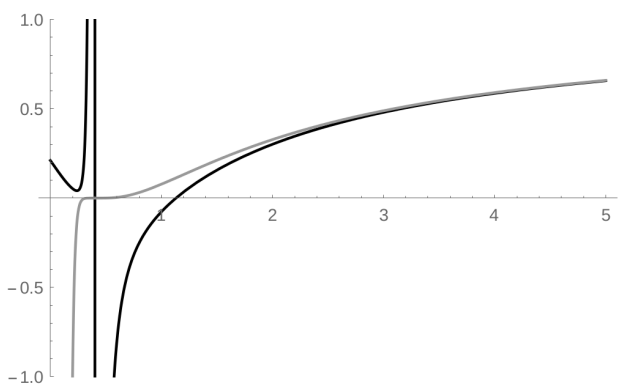}  
\caption{Behaviour of $g_{tt}(r)$ (Gray) and $g^{-1}_{rr}(r)$ (Black) for $k=4$.
We see a zero and a singular point for $g^{-1}_{rr}$. The singularity at $r_c=2/5$ is hidden behind the horizon $h\simeq 1.131$.
It can be seen that the black hole horizon is shifted inside the Schwarzschild radius
($r_s=2\,{\cal M}$) by extra-dimensional effects.
The ADM mass is ${\cal M}=1$.
 \label{BH}}
\end{center}
\end{figure}
\par
%
%\begin{figure}[pb]
%\begin{center}\includegraphics[width=3.2in]{UP}  
%\caption{Behaviour of the effective density ${\cal U}(r)$ (Black) and the anisotropy ${\cal P}(r)$ (Gray) for $k=4$.
%The scalar function ${\cal U}$ 
%increasing as we approach to the distribution until reach a maximun value inside the horizon $h$, then it diverges as we approach to the singularity $r_c$. 
%The opposite behaviour is seen for ${\cal P}$. The black hole mass is setting as ${\cal M}=1$.
% \label{UP}}
%\end{center}
%\end{figure}
%\par
%
\begin{figure}[pb]
\begin{center}\includegraphics[width=3.2in]{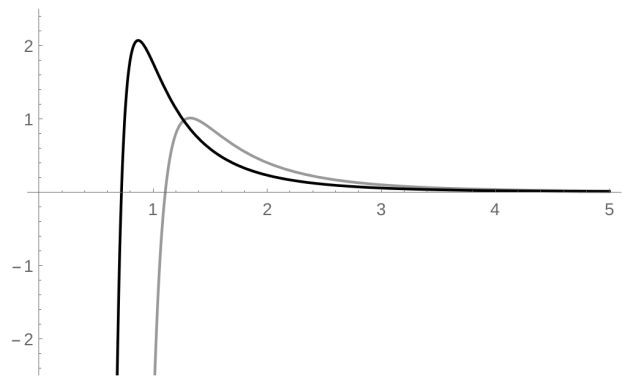}  
\caption{Comparison of the effective density ${\cal U}(r)$ for the case $k=2$ (gray) and $k=4$ (black). In both cases the scalar function ${\cal U}$ 
increasing as we approach to the distribution until reach a maximun value. This maximun is slightly beyond the horizon for $k=2$ and 
inside it for $k=4$. In both cases the function diverges as we approach to their respective singularities $r_c$. The black hole mass is setting as ${\cal M}=1$.
 \label{U}}
\end{center}
\end{figure}
\par
\begin{figure}[pb]
\begin{center}\includegraphics[width=3.2in]{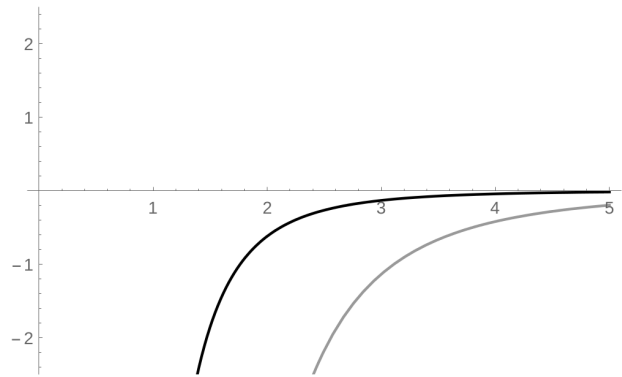}  
\caption{Comparison of the anisotropy ${\cal P}(r)$ for the case $k=2$ (gray) and $k=4$ (black). In both cases ${\cal P}$ is incremented negatively 
and diverges as we approach to their respective singularities $r_c$. The black hole mass is setting as ${\cal M}=1$.
 \label{P}}
\end{center}
\end{figure}
\begin{equation}
\label{deftemp}
e^{\nu}
=
1-\frac{2 {\cal M}}{r}+\frac{{Q}}{r^2}-\frac{2}{5}\frac{{\cal M}Q}{r^3} + \frac{{Q}^2}{20\,r^4}
-\frac{{\cal M} Q^2}{250\,r^5}\ ,
\end{equation}
where ${\cal M}=5\,M$ and $Q=40\,M^2$. The tidal charge $Q$ is nothing but extra-dimensional effects producing deviation from the Schwarzschild's solution. Similarly the 
radial metric component is expressed in an exact form, too large to display here. 
This solution displays a zero of $g^{-1}_{rr}$, namely a horizon $r=h$, and
a surface $r=r_c$ where $g^{-1}_{rr}$ diverges (and $g_{tt}=0$), all shown in Fig.~\ref{BH}. 
These surfaces separate the space-time in three regions, namely
\begin{itemize}
 \item $0 < r < r_c$
  \item $r_c < r < h$
 \item $r > h$.
 \end{itemize}
We want to emphasize that an exterior observer at $r > h$ will never see the singularity at $r=r_c$, as it is hidden behind 
the horizon $r=h$. Finally, since the horizon lies inside the Schwarzschild radius, $h<r_s=2\,{\cal M}$,
this solution clearly indicates that extra-dimensional effects weaken the gravitational field. The solution also describes the exterior of a stellar structure with 
radius $R>h$.
\par
Both Weyl functions ${\cal U}$ and ${\cal P}$ are shown in Fig.~\ref{U} and Fig.~\ref{P} respectively and compared with the case $k=2$. 
While the scalar ${\cal U}(r)$ is always positive, the Weyl functions ${\cal P}(r)$ is negative. 
This indicates a negative radial deformation and also a positive temporal deformation, according to Eqs.~\eqref{UU} and~\eqref{P}, and diverges at the 
singular surface $r=r_c$. Both functions represent the Weyl fluid around the self-gravitating system as consequences of extra-dimensional effects. 
It is interesting to note that in this new case with $k=4$ the Weyl fluid is weaker than the studied in Ref.~\cite{MGDextended} with $k=2$.

\par

\section{Conclusions}
\par

In the context of the BW, and by using the extension of the MGD approach, where both gravitational potentials are deformed by bulk gravitons, 
the exterior spacetime $r>R$ associated to a BW self-gravitating system of radius $R$ was studied, and the consequences of extra-dimensional effects 
on the Schwarzschild solution were investigated. By using the specific time deformation $h(r)$ in Eq.~\eqref{tdef}, a master equation showing the deformation undergone 
by the radial metric component was identified in Eq.~\eqref{master}. This master solution depends on three free parameters, the GR mass 
$M$ (ADM mass without bulk effects), the time parameter $k$, and an integration constant $\beta$ which is fixed 
demanding the Schwarzschild limit $g_{rr}^{-1}\sim\,1-2{\cal M}/r$, where the ADM mass ${\cal M}$ and the GM mass $M$ are related by ${\cal M}=(1+k)\,M$.

Taking a specific value for the time parameter, namely $k=4$, a new exact exterior solution for a 
spherically symmetric self-gravitating systems was identified. The main feature of this solution is the presence 
of higher-order terms in the tidal charge $Q$, thus generalizing the well known tidally charged black hole solution found in Ref.~\cite{qmetric}. 
This new solution, which has a singularity $r_c$ hidden behind the horizon $h$, represents a black hole 
whose horizon lies inside the Schwarzschild radius, $h<r_s=2\,{\cal M}$, indicating that 
extra-dimensional effects weaken the gravitational field. The scalar function ${\cal U}$ 
increasing as we approach to the distribution until reach a maximun value inside the horizon $h$, then it diverges as we approach to the singularity $r_c$. 
The opposite behaviour is seen for ${\cal P}$, which is always negative. Both functions tend to disappear rapidly as we move away from the stellar distribution, 
showing thus a ``Weyl atmosphere'' surrounding the self-gravitating system. The new solution was compared with the previous one obtained in 
Ref.~\cite{MGDextended}. Despite the fact that in this new solution the time deformation is greater than the one in Ref.~\cite{MGDextended}, 
the Weyl fluid surronding the self-gravitating system is weaker. This shows that the simplest way to introduce a deformation on the temporal metric component, namely 
the expression shown in Eq.~\eqref{g00def}, has non trivial consequences. In this respect, it would be interesting to use 
the classical tests of General Relativity in the Solar system, as done in Ref.~\cite{jo12}, to derive bounds on the time parameter $k$, as well as to investigate 
the extension of this solution in the complete five-dimensional bulk to clarify the role of $k$. Finally, despite the fact that in the braneworld 
it is not clear whether the process of gravitational collapse leaves a signature in the black hole end-state or not,\cite{bh1}-\cite{bh3} the study of the 
``no hair theorem'' by using these higher-order tidal charge solutions represents an attractive scenario.

\subsection*{Acknowledgments}
This work is supported by the Abdus Salam International Center for Theoretical Physics, ICTP, and partially for 
the European Union under Erasmus Mundus program, grant 2012-2646 / 001-001-EMA2. The author also thanks 
Dipartimento di Fisica e Astronomia, Alma Mater Universit\`a di Bologna, for kind hospitality.

\begin{thebibliography}{99}
%\footnotesize
%
\bibitem{Ber} 
E. Bergshoeff, O. Hohm and P. K. Townsend, { Phys. Rev. Lett.} {\bf 102}, 201301 (2009),  
{\it Massive Gravity in Three Dimensions}, arXiv:0901.1766. 

\bibitem{claudia} 
C. de Rham, { Living Rev. Relativity} {\bf 17}, 7 (2014), 
{\it Massive gravity}, arXiv:1401.4173v2 [hep-th].

\bibitem{galileon} 
N. Chow and J. Khoury { Phys. Rev. D} {\bf 80}, 024037 (2009), 
{\it Galileon Cosmology}, arXiv:0905.1325v4 [hep-th]. 

\bibitem{diFelice} 
A. De Felice and S. Tsujikawa,	{Living Rev. Rel.} {\bf 13}, 3 (2010),
{\it f(R) Theories}, arXiv:1002.4928v2 [gr-qc].

\bibitem{thomas} 
T. P. Sotiriou and V. Faraoni, { Rev. Mod. Phys.} {\bf 82}, 451 (2010),
{\it f(R) Theories Of Gravity}, arXiv:0805.1726v4 [gr-qc].

\bibitem{salvatore} 
S. Capozziello and M. De Laurentis, { Phys. Rept.} {\bf 509}, 167 (2011), 
{\it Extended Theories of Gravity}, arXiv:1108.6266v2 [gr-qc].

\bibitem{salvatore2} 	
S. Capozziello, V. F. Cardone and A. Troisi, { Phys. Rev. D} {\bf 71}, 043503 (2005), 
{\it Reconciling dark energy models with f(R) theories}, arXiv:astro-ph/0501426v1.

\bibitem{padilla} T. Clifton, P. G. Ferreira, A. Padilla and C. Skordis, {Phys. Rept.} {\bf 513}, 1 (2012),
{\it Modified Gravity and Cosmology}, arXiv:1106.2476v3 [astro-ph.CO]. 

\bibitem{horava} P. Horava, {Phys. Rev. D} {\bf 79}, 084008 (2009), 
{\it Quantum Gravity at a Lifshitz Point}, arXiv:0901.3775v2 [hep-th].
%
%
\bibitem{lisa1}
L.~Randall and R.~Sundrum,
Phys. Rev. Lett. {\bf 83}, 3370 (1999), {\it A Large Mass Hierarchy from a Small Extra Dimension,} arXiv:hep-ph/9905221v1.
%
\bibitem{lisa2}L.~Randall and R.~Sundrum,
Phys. Rev. Lett. {\bf 83}, 4690 (1999), {\it An Alternative to Compactification}, arXiv:hep-th/9906064.
%
%
\bibitem{ADD}
N. Arkani-Hamed, S. Dimopoulos, G. Dvali,
Phys.\,Lett.\,B {\bf 429} 263 (1998), {\it The Hierarchy Problem and New Dimensions at a Millimeter}, arXiv:hep-ph/9803315. 
\bibitem{ADD1}
I. Antoniadis, N. Arkani-Hamed, S. Dimopoulos, G. Dvali,
Phys. Lett. B {\bf 436}, 257 (1998), {\it New Dimensions at a Millimeter to a Fermi and Superstrings at a TeV}, arXiv:hep-ph/9804398.
%
%
\bibitem{laszlo2011}
L. \'A Gergely, T. Harko, M. Dwornik, G. Kupi, Z. Keresztes,
Mon. Not. Royal Astron. Soc. {\bf 415}, 3275 (2011), {\it Galactic rotation curves in brane world models},  
arXiv:1105.0159 [gr-qc].
%
%\bibitem{Gergely:2006hd}
%  L.~A.~Gergely,
%  \emph{Black holes and dark energy from gravitational collapse on the brane,}
%  \emph{JCAP} {\bf 0702 } (2007)  027
%  [{\tt  arXiv:hep-th/0603254}].
%
\bibitem{shi} T. Shiromizu, K. Maeda and M. Sasaki, { Phys. Rev. D} {\bf 62}, 024012 (2000),
{\it The Einstein equations on the 3-brane world}, arXiv:gr-qc/9910076v3. 

\bibitem{12} C. Germani and R. Maartens, Phys. Rev. D {\bf 64}, 124010 (2001), {\it Stars in the braneworld}, arXiv:hep-th/0107011.

\bibitem{7} 
  R.~da Rocha, J.~M.~Hoff da Silva,
  Phys.\ Rev.\ D {\bf 85}, 046009 (2012),  \emph{Black string corrections in variable tension braneworld scenarios},
  arXiv:1202.1256 [gr-qc].

\bibitem{13} R.~Casadio, A.~Fabbri and L.~Mazzacurati, Phys.\ Rev.\ D {\bf 65},  084040 (2002), {\it New black holes in the brane world?}, gr-qc/0111072.

\bibitem{14} P. Kanti and K. Tamvakis, { Phys. Rev. D} {\bf 65}, 084010 (2002), {\it Quest for localized 4D black holes in brane worlds}.

\bibitem{17} G. Kofinas, E. Papantonopoulos and V. Zamarias, { Phys. Rev. D} {\bf 66}, 104028 (2002), {\it Black Hole Solutions in Braneworlds with Induced Gravity}, 
arXiv:hep-th/0208207v2.

\bibitem{maia1} M. D. Maia, E. M. Monte, J.M.F. Maia, Phys. Lett. A {\bf 297}, 9 (2002), {\it Geometry of brane worlds}, astro-ph/0208223. 

\bibitem{maia2} M. D. Maia, E. M. Monte, J.M.F. Maia, Phys. Lett. B {\bf 585}, 11 (2004), {\it The Accelerating universe in brane world cosmology}, astro-ph/0208223.
 
\bibitem{18} 
G. Kofinas, E. Papantonopoulos, I. Pappa, Phys.Rev.D {\bf 66} 104014 (2002), {\it Spherically Symmetric Braneworld Solutions with $R_{4}$ term in the Bulk}, arXiv:hep-th/0112019.
%

\bibitem{8} L. A. Gergely, { Phys. Rev. D} {\bf 68}, 124011 (2003), {\it Generalized Friedmann branes}, arXiv:gr-qc/0308072v2.

\bibitem{9} L. A. Gergely, Phys. Rev. D {\bf  78},  084006 (2008), {\it Friedmann branes with variable tension}, arXiv: 0806.3857 [gr-qc].

\bibitem{5} R. Casadio and L. Mazzacurati, { Mod. Phys. Lett. A} {\bf 18}, 651 (2003), {\it Bulk shape of brane-world black holes}, arXiv:gr-qc/0205129v2. 

\bibitem{22} P. Kanti, I. Olasagasti and K. Tamvakis, {Phys. Rev. D} {\bf 68}, 124001 (2003), 
{\it Quest for localized 4D black holes in brane worlds. II. Removing the bulk singularities}, arXiv:hep-th/0307201v2.

\bibitem{19} M. Visser and D. L. Wiltshire, {Phys. Rev. D} {\bf 67}, 104004 (2003), {\it On-brane data for braneworld stars}, arXiv:hep-th/0212333v2.

\bibitem{20} A. S. Majumdar and N. Mukherjee, {Int. J. Mod. Phys. D} {\bf 14}, 1095 (2005), {\it 
Braneworld black holes in cosmology and astrophysics}, arXiv:astro-ph/0503473v2. 

\bibitem{2} S. Creek, R. Gregory, P. Kanti, B. Mistry, Class. Quant. Grav. {\bf 23}, 6633 (2006), {\it Braneworld stars and black holes}, arXiv:hep-th/0606006.
%

\bibitem{11} L. A. Gergely, {J. Cosmol. Astropart. Phys.} {\bf 02}, 027 (2007), {\it 
Black holes and dark energy from gravitational collapse on the brane}, arXiv:hep-th/0603254v3. 

\bibitem{6} R. Casadio and O. Micu, {Phys. Rev. D} {\bf 81}, 104024 (2010), {\it Exploring the bulk of tidal charged micro-black holes}, arXiv:1002.1219v2 [hep-th].

\bibitem{1} R. Maartens and K. Koyama, {Living Rev. Rel.} {\bf 13}, 5 (2010), {\it Brane-World Gravity}, arXiv:1004.3962v2 [hep-th].

\bibitem{3} D. C. Dai and D. Stojkovic, {Phys. Lett. B} {\bf 704}, 354 (2011), {\it Analytic solution for a static black hole in the RSII model}, 
arXiv:1004.3291v2 [gr-qc].

\bibitem{16} R. da Rocha and J.M. Hoff da Silva, {Phys. Rev. D} {\bf 85}, 046009 (2012), {\it Black string corrections in variable tension braneworld scenarios}, 
arXiv:1202.1256v1 [gr-qc]. 

\bibitem{4} S. Abdolrahimi, C. Cattoen, D. N. Page and S. Yaghoobpour-Tari, {Phys. Lett. B} {\bf 720}, 405 (2013), {\it Large Randall-Sundrum II Black Holes}, 
arXiv:1206.0708v4 [hep-th]. 

\bibitem{15} D. Bazeia, J. M. Hoff da Silva and R. da Rocha, {Phys. Rev. D} {\bf 90}, 047902 (2014), 
{\it Regular Bulk Solutions and Black Strings from Dynamical Braneworlds with Variable Tension}, arXiv:1401.6985v2 [hep-th]. 


\bibitem{25} L. B. Castro, M. D. Alloy and D. P. Menezes, {JCAP} {\bf 1408},  047 (2014), {\it Mass radius relation of compact stars in the
braneworld}, arXiv:1403.1099v2 [nucl-th].

\bibitem{26} T. Harko and M. J. Lake, {Phys. Rev. D} {\bf 89}, 064038 (2014), {\it Null fluid collapse in brane world models}, arXiv:1312.1420v3 [gr-qc]. 

\bibitem{24} S. Chakraborty and S. SenGupta, {Eur. Phys. J C} {\bf 75}, 11 (2015), {\it Spherically symmetric brane spacetime with bulk $f(\mathcal{R})$ gravity}, 
arXiv:1409.4115v2 [gr-qc]. 

\bibitem{23} F. X. Linares, M. A. Garcia-Aspeitia and L. A. Ureña-Lopez, {Phys. Rev. D} {\bf 92}, 024037 (2015), {\it Stellar models in Brane Worlds}, 
arXiv:1501.04869v1[gr-qc].

\bibitem{kantix} P. Kanti, N. Pappas and T. Pappas, Class. Quant. Grav. {\bf 33}, 015003 (2016).
{\it On the Localisation of 4-Dimensional Brane-World Black Holes II: the general case}, 
arXiv:1507.02625v1[hep-th]. 
%%%%%%%%%%%%%%%%%%%%%%%%%%%%%%%%%%%%%%%%%%%%%%%%%%%
\bibitem{Abdalla} M. C. B. Abdalla, P. F. Carlesso and J. M. Hoff da Silva, Eur. Phys. J. C {\bf 75}, 432 (2015), 
{\it Solution for a local straight cosmic string in the braneworld gravity}, arXiv:1507.03968v2 [gr-qc].

\bibitem{Miguel} Miguel A. Garcia-Aspeitia, Eur. Phys. J. C {\bf 75}, 530 (2015), {\it Branes constrictions with White Dwarfs}, arXiv:1510.06814v1 [gr-qc].

\bibitem{Miguel2} Miguel A. Garcia-Aspeitia and  Mario A. Rodriguez-Meza, {\it Constraining brane tension using rotation curves of galaxies}, arXiv:1509.05960v2 [gr-qc]. 

\bibitem{Miguel3} Miguel A. Garcia-Aspeitia, Mayra J. Reyes-Ibarra, C. Ortiz, J. C. Lopez-Dominguez and Sinhue Hinojosa-Ruiz, 
{\it Gravitational collapse in brane-worlds: the dynamical systems perspective}, arXiv:1412.3496v1 [gr-qc].

\bibitem{Banerjee} Ayan Banerjee, Farook Rahaman, Sayeedul Islam and Megan Govender, {\it Braneworld gravastars admitting conformal motion}, arXiv:1510.05939v2 [gr-qc].

\bibitem{Fran} Francisco Linares, Miguel A. Garcia-Aspeitia and L.Arturo Urena-Lopez, J. Phys. Conf. Ser. {\bf 545}, 012007 (2014), {\it Stellar Stability in the Braneworld}.

\bibitem{Sumanta} Sumanta Chakraborty and Soumitra SenGupta, Eur. Phys. J C {\bf 75}, 11 (2015), {\it Spherically symmetric brane spacetime with bulk f(R) gravity}, arXiv:1409.4115v2 [gr-qc]. 

\bibitem{Pappa2} Nikolaos D. Pappas, {\it The black hole challenge in Randall-Sundrum II model}, arXiv:1409.0817v1 [gr-qc]. 

\bibitem{Gonzalez} R. Gonzalez Felipe, D. Manreza Paret and A. Perez Martinez, {\it Constraints on the braneworld from compact stars}, arXiv:1601.01973v1 [gr-qc].

\bibitem{RRR} Roberto Casadio, Rogerio T. Cavalcanti and Roldao da Rocha, {\it Fluid/gravity correspondence and the CFM brane-world solutions}, arXiv:1601.03222v1 [hep-th]. 
%%%%%%%%%%%%%%%%%%%%%%%%%%%%%%%%%%%%%%%%%%%%%%%%%%%%%
\bibitem{jo1} 
J. Ovalle,
Braneworld stars: anisotropy minimally projected onto the brane, {\it in Gravitation and Astrophysics}
(ICGA9), ed. J. Luo. (World Scientific, Singapore, 2010), p. 173., arXiv:0909.0531v2 [gr-qc].

\bibitem{GD100} 
Jorge Ovalle, Beyond Einstein Gravity: The Geometric Deformation, 
{\it General relativity, 100 years after Hilbert}, edited by Jan Brajercik,
ISBN 978-80-555-13300, Presov, Slovakia. University of Presov Editorial, 2015.
p 34-37.

\bibitem{jo2} 
J. Ovalle, {Mod. Phys. Lett. A}, {\bf 23}, 3247 (2008), {Searching Exact Solutions for Compact Stars in Braneworld: a conjecture}, arXiv:gr-qc/0703095v3.

\bibitem{jo3} 
J.  Ovalle, {Int.  J.  Mod.  Phys.  D}, {\bf 18},  837  (2009), 
{\it Non-uniform  Braneworld  Stars:  an  Exact  Solution}, arXiv:0809.3547 [gr-qc].

\bibitem{jo4} 
J. Ovalle, {Mod. Phys. Lett. A}, {\bf 25}, 3323 (2010), 
{\it The Schwarzschild's Braneworld Solution}, arXiv:1009.3674 [gr-qc].

\bibitem{jo5}
J. Ovalle, {\it Effects of density gradients on braneworld stars}, in {\it 
Proceedings of the Twelfth Marcel Grossmann Meeting on General Relativity}, eds. Thibault Damour, Robert T. Jantzen and Remo Ruffini. 
(World Scientific, Singapore, 2012), p. 2243. 

\bibitem{qmetric} N. Dadhich, R. Maartens, P. Papadopoulos, and V. Rezania, {Phys. Lett. B} {\bf 487}, 1 (2000), 
{\it Black  holes  on  the  brane}, hep-th/0003061.

\bibitem{jo6} 
R.  Casadio and J. Ovalle, {Phys.  Lett.  B}, {\bf 715}, 251
(2012), {\it Brane-world stars and (microscopic) black holes}, arXiv:1201.6145 [gr-qc].

\bibitem{jo7} 
R. Casadio and J. Ovalle, {Gen. Relat. Grav.} {\bf 46}, 1669 (2014), 
{\it Brane-world stars from minimal geometric deformation, and black holes}, arXiv:1212.0409v2 [gr-qc].

\bibitem{jo8} 
J. Ovalle and F. Linares, {Phys. Rev. D}, {\bf 88}, 104026 (2013), {\it Tolman IV solution in the Randall-Sundrum Braneworld}, arXiv:1311.1844v1 [gr-qc].

\bibitem{jo9} 
J. Ovalle, F. Linares, A Pasqua and A Sotomayor, {Class. Quantum Grav.} {\bf 30}, 175019 (2013), {\it The role of exterior Weyl fluids on compact stellar structures in Randall-Sundrum gravity}, arXiv:1304.5995v2 [gr-qc].

%
\bibitem{ger1}
L. A. Gergely, {Phys. Rev. D} {\bf  78}, 084006 (2008), {\it Friedmann branes with variable tension}, arXiv: 0806.3857 [gr-qc].
% [{\tt arXiv: 0806.3857v3 [gr-qc]}].

\bibitem{jo10} 
 R. Casadio, J. Ovalle and R. da Rocha, {Class. Quantum
Grav.} {\bf 30}, 175019 (2014), {\it Black Strings from Minimal Geometric Deformation in a Variable Tension Brane-
World}, arXiv:1310.5853 [gr-qc].

\bibitem{jo11} 
J. Ovalle, L. A. Gergely and R. Casadio, {Class. Quantum
Grav.} {\bf 32}, 045015 (2015), {\it Brane-world stars with solid crust and vacuum exterior}, arXiv:1405.0252v2 [gr-qc].

\bibitem{jo12}  R. Casadio, J. Ovalle and R. da Rocha, {Europhys. Lett.} {\bf 110}, 40003 (2015), 
{\it Classical Tests of General Relativity: Brane-World Sun from Minimal Geometric Deformation}, arXiv:1503.02316 [gr-qc].

\bibitem{MGDextended} R. Casadio, J. Ovalle and R. da Rocha, Class. Quant. Grav. {\bf 32}, 215020 (2015),
{\it The Minimal Geometric Deformation Approach Extended},  arXiv:1503.02873v2 [gr-qc]. 

\bibitem{tiberio1} M. K. Mak and T. Harko, {Phys. Rev. D} {\bf 70}, 024010 (2004), {\it Can the galactic rotation curves be explained in brane world models?}, 
arXiv:gr-qc/0404104v1. 

\bibitem{tiberio2} C. G. Boehmer and T. Harko, {Class. Quantum. Grav.} {\bf 24}, 3191 (2007), {\it Galactic dark matter as a bulk effect on the brane}, 
arXiv:0705.2496v3[gr-qc]. 

\bibitem{tiberio3} L. A. Gergely, T. Harko, M. Dwornik, G. Kupi and Z. Keresztes, {Mon. Not. R. Astron. Soc.} {\bf 415}, 3275 (2011), 
{\it Galactic rotation curves in brane world models}, arXiv:1105.0159v2[gr-qc]. 

\end{thebibliography}
\end{document}